\documentclass{article}

\usepackage{arxiv}

\usepackage[utf8]{inputenc} 
\usepackage[T1]{fontenc}    
\usepackage{hyperref}       
\usepackage{url}            
\usepackage{booktabs}       
\usepackage{amsfonts}       
\usepackage{nicefrac}       
\usepackage{microtype}      
\usepackage{lipsum}
\usepackage{graphicx}
\usepackage{subcaption}
\usepackage{amsmath}
\usepackage{amsthm}

\title{Emotion Manipulation Through Music - A Deep Learning Interactive Visual Approach}

\author{
  Adel N. Abdalla\\
  Data Science Institute \\
  Columbia University, New York, NY, USA\\
\texttt{ana2169@columbia.edu} \\
    \And
   Jared Osborne\\ 
  Computer Science Department\\
  Central Washington University, 
  Ellensburg, WA, USA\\
    \texttt{osbornejar@cwu.edu} \\\\
    \And
   R\u azvan Andonie\\
   Computer Science Department\\
   Central Washington University, Ellensburg, WA, USA\\
   and\\
   Transilvania University of Bra\c sov, Romania\\
   \texttt{razvan.andonie@cwu.edu}
}

\begin{document}

\maketitle

\begin{abstract}
Music evokes emotion in many people. We introduce a novel way to manipulate the emotional content of a song using AI tools. Our goal is to achieve the desired emotion while leaving the original melody as intact as possible. For this, we create an interactive pipeline capable of shifting an input song into a diametrically opposed emotion and visualize this result through Russel's Circumplex model. Our approach is a proof-of-concept for Semantic Manipulation of Music, a novel field aimed at modifying the emotional content of existing music. We design a deep learning model able to assess the accuracy of our modifications to key, SoundFont instrumentation, and other musical features. The accuracy of our model is in-line with the current state of the art techniques on the 4Q Emotion dataset. With further refinement, this research may contribute to on-demand custom music generation, the automated remixing of existing work, and music playlists tuned for emotional progression.
\end{abstract}


\section{Introduction} \label{sec:introduction}
In recent years, the fields of Music Information Retrieval (MIR) and Music Emotion Recognition (MER) have received significant attention, leading to multiple advances in how music is analyzed \cite{Han2022, INR-042}. These developments have increased the accuracy in determining what emotions are present in a given music sample, but the current state of the art is only now passing 75\% through the use of Random Forest and Support Vector Machine models \cite{10100058}. This is in contrast to the field of speech recognition, where current models are approaching 100\% accuracy across hundreds of languages for word identification \cite{zhang2023google} and 85\% for standard speech emotion recognition \cite{kumar2023multilayer}. 

The additional challenges in music recognition come from the nature of music itself as the lyrical and emotional content of a vocalist's contribution are only one part of the whole. Tempo, rhythm, timbre, instrumentation choice, perceived genre, and other factors contribute together to shape the emotional and tonal landscape of any given work into a unique blend that is interpreted subjectively by individual listeners \cite{thompson2023psychological}.

The goal of our paper is to show that by changing the underlying structure of a small subset of musical features of any given musical piece, we can adjust the perceived emotional content of the work towards a specific desired emotion. This appears to be a novel approach, as current research into MIR and MER is focused on improving the accuracy of existing emotional content identification rather than changing it to a new emotion. Our approach fits into the visual and deep learning framework. In other words, transformations are applied to the audio, a deep learning model is applied, and visuals are produced. This makes it very different than the interactive control system for emotional expression in music presented in \cite{Grimaud2021}, which allows users to make changes to both structural and expressive cues (tempo, pitch, dynamics, articulation, brightness, and mode) of music in real-time.

The contributions in this paper are threefold:
\begin{enumerate}
    \item We create a pipeline capable of shifting an input song into a diametrically opposed emotion and visualize this result through Russel's Circumplex model \cite{Russel1980}.  
    \item We generate a classifier that is able to map a given song into an emotional quadrant of Russel's Circumplex model.
    \item We build the groundwork for the novel field of what we call Semantic Manipulation of Music (SMM).
\end{enumerate}

Our work can be accessed and replicated through the provided  Github\footnote{\url{https://github.com/aa221/Semantic-Manipulation-of-Music}} code.

The visual aspect of our framework draws on Russell's Circumplex model \cite{Posner2005-sw}, which plots emotional context on a 2D plane where the x-axis represents arousal (or excitement) and the y-axis represents valence (or pleasure). It has been used in multiple fields to quantify emotion, such as representing the physical motion of a motion-captured dancer performing to music as emotional information \cite{10.1525/mp.2013.30.5.517} or in MER and MIR to determine the overall mood of a given audio sample.

Section \ref{sec:previous} of the paper summarizes previous results related to music emotion prediction, including the software libraries we use in our approach. In Section \ref{sec:ourapproach} we introduce the concept of SMM. Section \ref{sec:results} presents experimental results, whereas Section \ref{sec:conclusions} includes our final remarks and future work.

\section{Previous Work}\label{sec:previous}
This section summarizes previous work related to music sentiment analysis using deep learning and lists the techniques and software used by us to build our pipeline. 

\subsection{Emotion analysis using deep learning}
Emotion is a complicated and multi-facted notion in music and it is hard to objectively capture. Recent studies have utilized deep neural networks to extract and predict emotional information based on the semantics of the acoustic features in an audio sample. A large variety of neural architectures were used, including Convolutional Neural Networks, Recurrent Neural Networks, Multilayer Perceptron, Gated Recurrent Units, and Long Short Term Memory \cite{Cheuk2020, Dong2019, Liu2019, Thao2019, Mate2022, Jia2022, Tian2023, Nguyen2023}.

Several deep audio embedding methods were proposed to be used in the MER task. Deep audio embeddings are a type of audio features extracted by a neural network that take audio data as an input and compute features of the input audio \cite{Koh2021}. For instance, pre-trained networks like $L^{3}$-Net and VGGish with deep audio embeddings were used to predict emotion semantics in music without expert human engineering \cite{Koh2021}. Some studies also include sentiment analysis based on the lyrics of songs \cite{Jia2022}, EEG-based emotion classification \cite{Ahmed2022}, or emotion classification of music videos \cite{Ahmed2022}, but all these are beyond our current focus.

The paper of Ahmed \emph{et al.} \cite{Ahmed2022} focuses on how the perception of emotion in music is affected by the timbre of the instruments used. Timbre is composed of overtones (higher-frequency standing waves) and harmonic series (higher-frequency tones that are integral multiples of the fundamental frequency) and these create a unique sound for every instrument. For each song, the emotion present was created through the balancing of instrumentation, melody, and other audio features by the original author. Therefore, Ahmed \emph{et al.} investigated whether separating each instrument’s part of a song into multiple individual waveforms could help improve the performance of models understanding emotion in music. In our work, we use the selection of instrumentation as a hyper-parameter to give us better control over the resultant emotions of our output. As each instrument has a unique timbre, the instrumentation itself shifts the semantic content of music towards or away from a target emotional content. 

Closer to our approach is the work of Ferreira \emph{et al.} \cite{Ferreira2019}, where the authors present a generative deep learning model that can be directed to compose music with a given sentiment. A major difference between our approach and \cite{Ferreira2019} is that we have two additional constraints: We require audio input as a starting point, and preserve as much of that original audio as possible so it can be recognized by the listener after a targeted sentiment manipulation has been performed. This imposes additional constraints by limiting the manipulation of the original melody to states where the original is still recognizable, considerably narrowing the range of possible outputs for our model and making the solution space smaller.

\subsection{Music21—transposing music}\label{sec:Music21—transposing music}

Music21 is an industry-standard toolkit\footnote{\url{https://web.mit.edu/music21/}} supported by M.I.T's School of Humanities, Arts, and Social Sciences and their Music and Theater Arts section, along with and generous grants from the Seaver Institute and the NEH/Digging-Into-Data Challenge. For our purposes, we are using the transposition feature to modify an input MIDI track to a different musical key.

\subsection{Accomontage2}\label{sec:Accomontage2}

AccoMontage2 is a toolkit\footnote{\url{https://github.com/billyblu2000/AccoMontage2}} capable of doing full-length song harmonization and accompaniment arrangement based on a lead melody \cite{zhao2021accomontage}. It can take a single melody as input and use a four-stage model to generate an accompanying audio track: a deterministic phrase-matching library that identifies note sequences inside of a musical bar and proposes an accompanying phrase, a fitness model that evaluates the appropriateness of the proposed phrase by evaluating the rhythm and chord features of the original and new phrase together, and a convolutional neural network to extract and evaluate features in the new and original phrases to create natural transitions. 

In our testing, we found that Accomontage produces a semi-deterministic result based upon the MIDI-formatted audio provided as input. By transposing the music to a different key before input, we manipulate the note-level phrases Accomontage2 matches against which changes the accompaniment generated and the overall mood of the musical work. In other words, we hypothesize that the correlation between the emotional content in a given key would be reflected by Accomontage2's selection of accompaniment, leading to the desired shift in emotional content in the resultant output.

\section{Semantic Manipulation of Music via Deep Learning}\label{sec:ourapproach}

\begin{figure}
    \centering
    \includegraphics[width=.7 \linewidth]{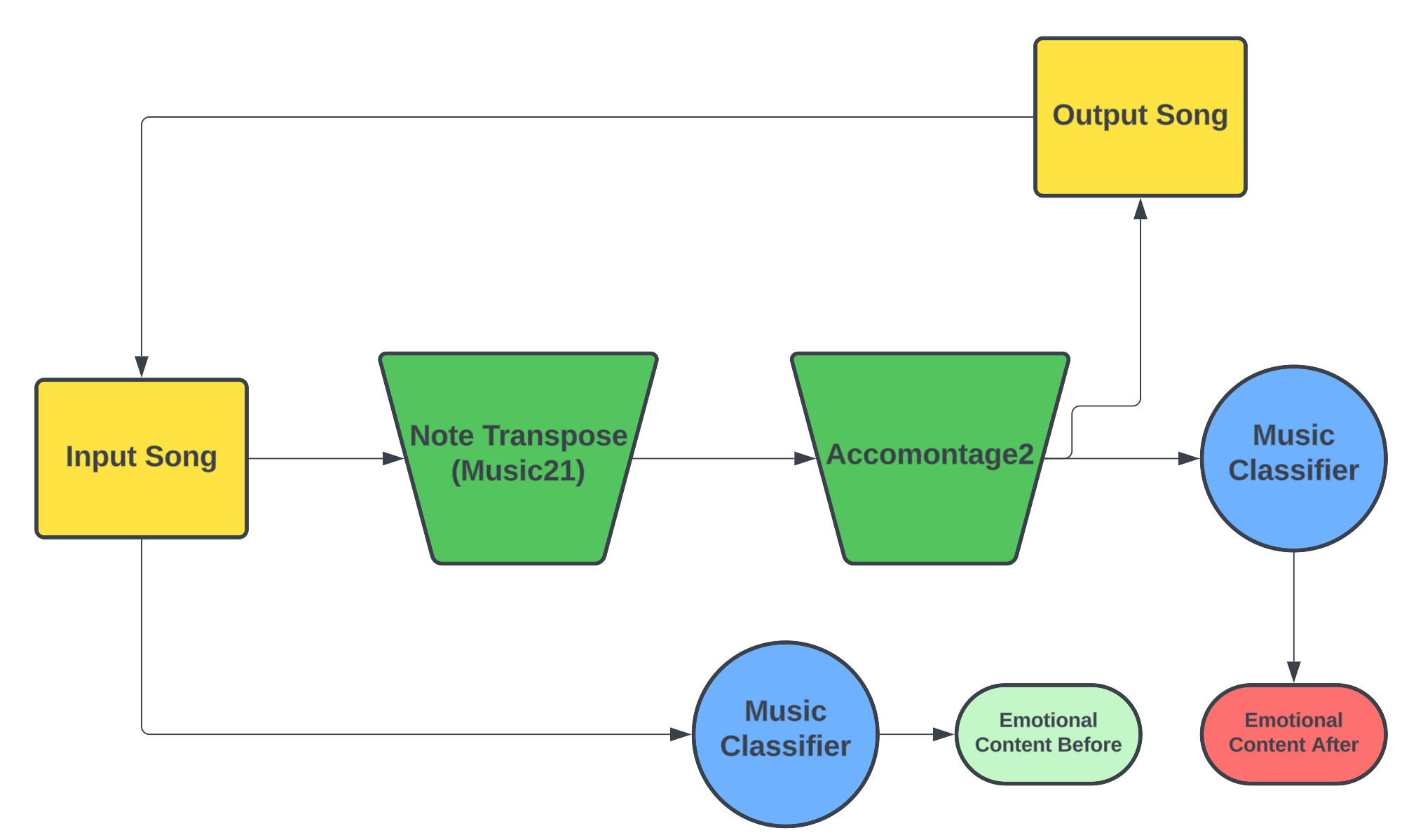}
    \caption{Pipeline Architecture.}
    \label{fig:pipeline_diagram}
\end{figure}

In this section we provide an overview of our pipeline architecture and classifier. 

Using a melody in MIDI format as input, we first synthesize the source audio and analyze its emotional content to serve as a baseline. We then chromatically transpose the MIDI file to a different key and pass this result through Accomontage2 to generate an accompanying track. After synthesizing the transposed and accompanied MIDI track, we evaluate the emotional content present to determine how much of a shift has occurred from the baseline. An example can be seen in Figs. 2 and 4-6. The transposition and accompaniment steps are repeated for multiple keys above and below the original audio's key, which allows us to select the result closest to our desired semantic result. We can visualize this process on Russell's Circumplex model in real-time, allowing us to visualize each result as its generated. This process can be seen in Fig. \ref{fig:pipeline_diagram}.

\subsection{Our Classifier}\label{sec:Our Classifier}

We use the Wav2Vec2Processor\cite{baevski2020wav2vec}, a processor with the ability to represent audio in a way that is understandable by our classifier. After using this processor we then feed the tokenized audio into a XLSR-Wav2Vec2 classification model. The original intent of this architecture was for speech transcription. In our current version, our toolchain works with instrumental music so the instrumentation is treated as speech in our analysis. Due to our limited compute resources, we selected sampling at the rate of 16KHz.

\begin{enumerate}
    \item Q1: represents happy emotions.
    \item Q2: represents angry emotions.
    \item Q3: represents sad emotions.
    \item Q4: represents calm emotions.
\end{enumerate}

Our work currently focuses on manipulating music from Q1 (happy and exciting) to Q3 (sad and relaxed) and vice versa as this represents the largest and most notable shift in both valence and arousal, though our approach is capable of moving a song towards any given quadrant.

\subsubsection{The architecture of the classifier}

Our model follows the architecture of the XLSR-Wav2Vec2 model  \cite{conneau2020unsupervised}. 

XLSR stands for "cross-lingual speech representations" and refers to XLSR-Wav2Vec2`s ability to learn speech representations that are useful across multiple languages. Like Wav2Vec2, XLSR-Wav2Vec2 learns speech representations from hundreds of thousands of hours of speech in more than 50 languages of unlabeled speech. With a character error rate of 2.87\% on the CommonVoice Test, the architecture of the model is adept enough to perform speech-to-text operations with extremely high accuracy. At a high level, the model contains convolution layers to understand the audio inputs, attention mechanisms to focus on the important portions of said inputs, and feed forward layers that allow for inference.

\subsubsection{Re-purposing the Model to Fit our Task}

To adapt the model for MER, we retrained the model on the 4Q emotional data set \cite{Panda2018}. In order to transform the model into a classification task, a classification head is applied to the model using the data set's .wav files and their corresponding quadrant labels. For a given song, our model returns four probabilities with a sum of 1.0, where each probability represents the song's likelihood of belonging to each corresponding quadrant. For example, an output of [.1, .6, .2, .1] indicates that the given song has a respective probability belonging to Q1, Q2, Q3 and Q4 of .1, .6, .2, .1.  

\subsection{Visualizing Results}

\begin{figure}[ht]
    \centering
    \includegraphics[width=0.4\linewidth]{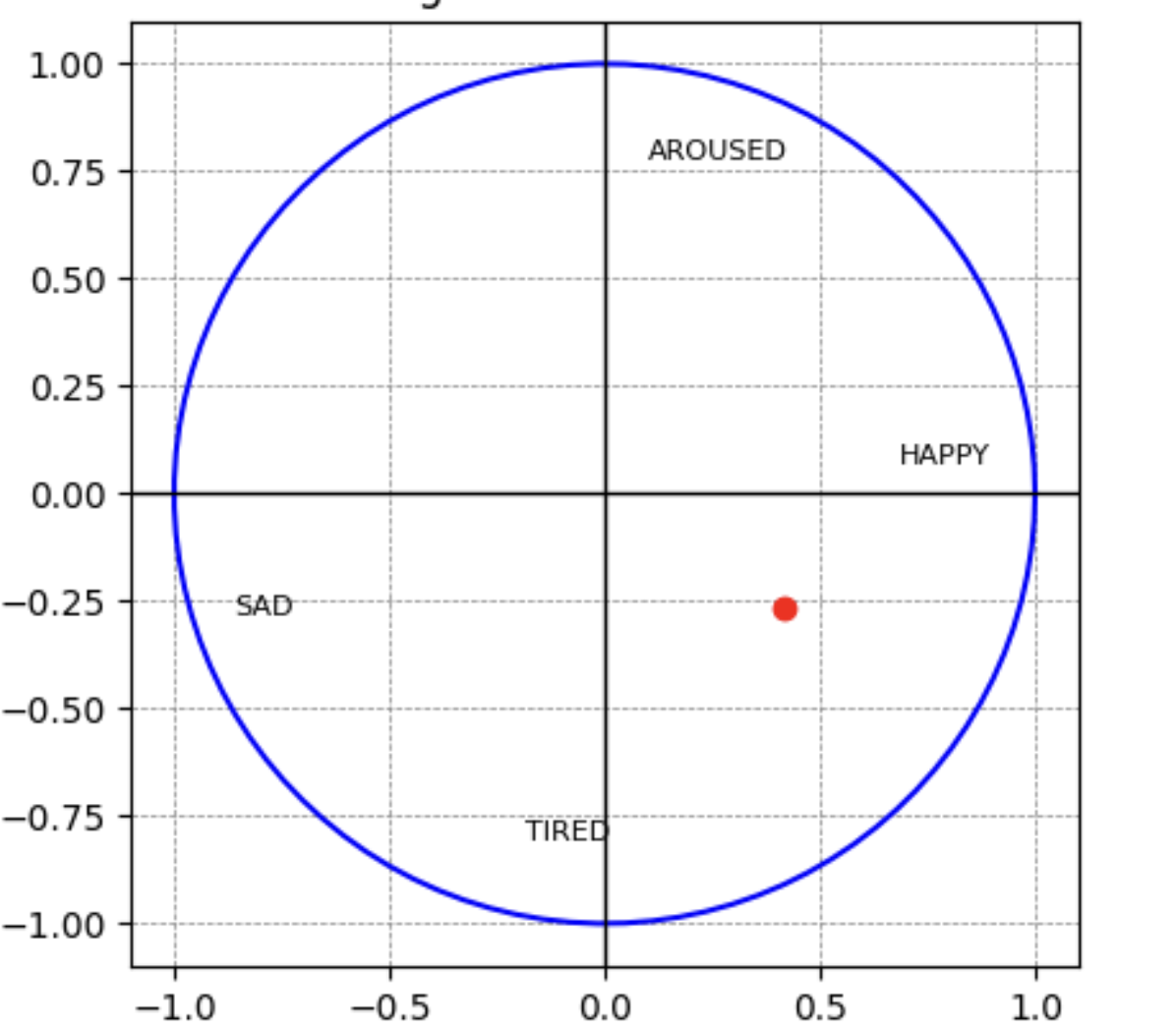}
    \caption{Visualization of song on our Russel Diagram.}
    \label{fig:circle_visualization}
\end{figure}

We employ Russel's Emotional Circumplex model \cite{Russel1980} (Fig. \ref{fig:circle_visualization}) to visualize the emotional content of each song. The X Axis corresponds to valence, the positive or negative degree of emotion, while the Y Axis corresponds to arousal, or the excitement level of the emotion. For example, a result of maximal arousal and valence would represent an energetic and happy song while one with minimal arousal and valence would represent a slow and sad song. 

Using our classifier output of a vector of four quadrant probabilities, \([p1, p2, p3, p4]\) we convert the probabilities into a \((x, y)\) coordinate: $x = (p1 - p3) \times \text{radius}$, $y = (p2 - p4) \times \text{radius}$. To ensure the point \((x, y)\) lies within the circle, we calculate its distance from the origin: $d = \sqrt{x^2 + y^2}$. If \(d > r\), the coordinates are normalized:
\[x_{\text{new}} = \frac{x}{d} \times r\qquad y_{\text{new}} = \frac{y}{d} \times r\] This allows us to plot a representative point on the Russel diagram while preserving the relative weight of each probability. An example of this may be seen in Fig \ref{fig:circle_visualization}.

\section{Experiments} \label{sec:results}

We have divided this section into two parts:
\begin{itemize}
    \item The prediction accuracy of our classifier vs. reported state of the art results. 
    \item A qualitative analysis on the effectiveness of our pipeline with respect to the visual output.
\end{itemize}

\subsection{Prediction accuracy}
For accurate comparison between our classifier and others, all models compared in this paper use the 4Q Audio Dataset\footnote{\url{https://www.kaggle.com/datasets/imsparsh/4q-audio-emotion-dataset-russell}}. This dataset quantifies the emotional content of 900 ~30 second clips of music into the four Circumplex quadrants, and this is what we use as the training and validation data for generating the classifier.

\begin{figure}
    \centering
    \includegraphics[width=.6 \linewidth]{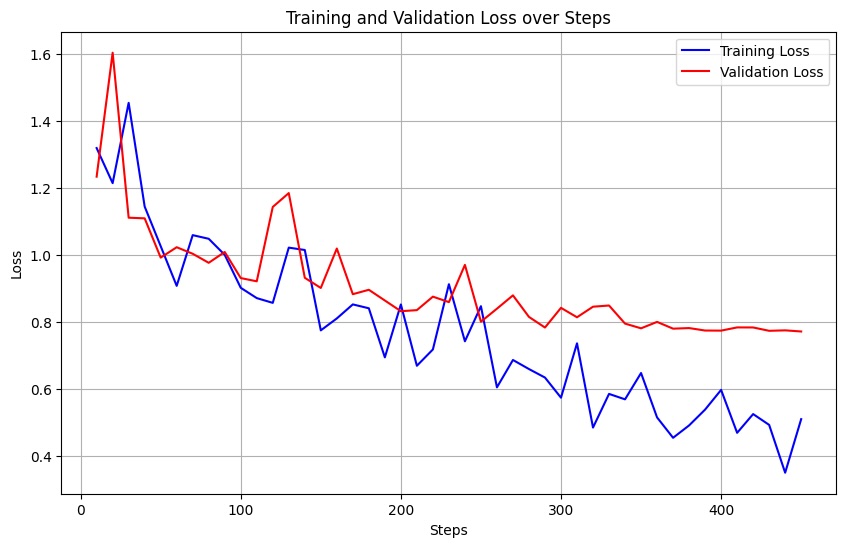}
    \caption{Training and Validation Loss for 4Q Dataset.}
    \label{fig:training_validation} 
\end{figure}

In our experiments, we used 10 training epochs, with 8 evaluation and train batch size per device. The learning rate was 1e-4.  We include Fig. \ref{fig:training_validation} to show the training and validation loss across all training steps. The decrease across steps indicates the model is learning and the lack of increase in validation loss indicates no over-fitting on the training set. 

With limited resources, our classifier was able to achieve an accuracy of 70\%, only 5\% lower than the accuracy of 75\% by Panda \emph{et al.} \cite{Panda2018} on their SVM model on the 4Q Emotional Dataset.  Table \ref{Table1} shows our classifier's performance relative to comparable models on the same dataset.

It should be noted that achieving the highest accuracy is not the purpose of this paper, as the classifier is an interchangeable component in our approach. As more accurate classifiers are developed, there should be a corresponding increase in manipulation efficiency in the field of SMM. Recently, Taiwanese authors have introduced additional input information from wearable devices measuring physiological data to increase accuracy, which resulted in a much higher accuracy of 92\% \cite{liao2022music}, but these results are not directly comparable with those in Table \ref{Table1}.

\begin{figure}
    \centering
    \includegraphics[width=.6 \linewidth]{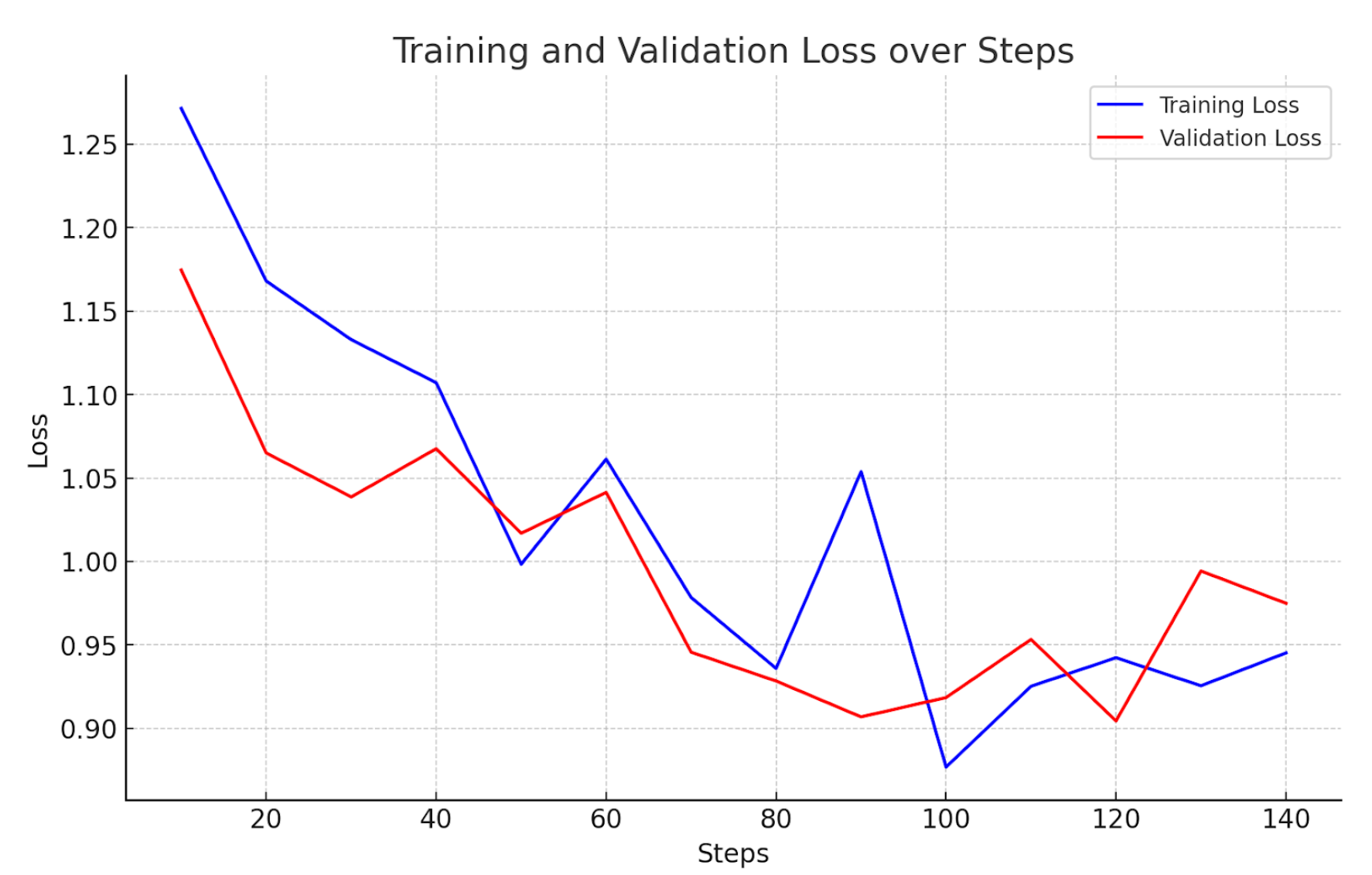}
    \caption{Training and Validation Loss for DEAM Dataset.}
    \label{fig:training_validation2} 
\end{figure}

\begin{table}[htbp]
\centering
\caption{Emotion Prediction Accuracy on the 4Q Audio Dataset.}
\footnotesize
\label{Table1}
\begin{tabular}{@{}lll@{}}
\toprule
Classifier & Accuracy \\ \midrule
Chaudhary \emph{et al.} \cite{Chaudhary2021} & 88.47\% &  \\
SVM \cite{Panda2018} & 73.5\% \\
L3-Net \cite{Koh2021} & 72\%  \\
\textbf{Our Classifier} & \textbf{70}\%  \\
MLP \cite{Pandrea2020} & 63\%  \\ \bottomrule
\end{tabular}
\end{table}

In order to fully assess the accuracy of the Classifier we also applied it to the DEAM (Database for Emotional Analysis using Music) dataset \cite{10.1371/journal.pone.0173392}. This dataset consists of 1902 excerpts and full songs annotated with valence and arousal values. Because the dataset did not contain Q1,Q2,Q3,Q4 values as the 4Q dataset did, the arousal and valence thresholds that were defined in the 4Q dataset were applied to the DEAM one. In this way, labels were engineered for the dataset. Applying the classifier to this dataset led to a 68\% classification accuracy after about 2 epochs (as opposed to the 10 epochs ran for the 4Q dataset). This performs relatively better than the classifier's performance on the 4Q dataset, as at 2 epochs in the previous experiment, the classifier had only achieved an accuracy of 68\%. Note the learning rate and batch size per device were the same as the 4Q experiment. Also note only 2 epochs were run due to a lack of GPU resource. The result on a completely separate dataset proves the robustness of the created classifier. Table \ref{Table2} reflects the accuracy of various classifiers on the DEAM data set. As one can see, the accuracy values are in line with the performance of our classifier.

\begin{table}[htbp]
\centering
\caption{Emotion Prediction Accuracy on the DEAM dataset.}
\footnotesize
\label{Table2}
\begin{tabular}{@{}lll@{}}
\toprule
Classifier & Accuracy \\ \midrule
Random Forest\cite{Medina} & 83\% &  \\
CNN \cite{Choi}& 80\% \\
\textbf{Our Classifier} & \textbf{68}\%  \\
MMD-MII Model\cite{Wang}  & 50\%  \\ \bottomrule
\end{tabular}
\end{table}

\subsection{Qualitative Analysis}
Our qualitative analysis focuses on the shift in emotional content as visualized in the Circumplex model. The question is if we have the ability to capture the emotional transformation of a song after its manipulation visually. This entails comparing the semantic content of a given song, before and after its transformations. We select the transformation closest to the desired result, with a successful transformation showing minimal deviation from the target result. This process is automated, as only the input melody and the target values for valence and arousal in the Circumplex model are required to evaluate the nearest result. 

\begin{figure*}[ht]
    \centering
    \begin{minipage}[b]{0.4\linewidth}
        \includegraphics[width=\linewidth]{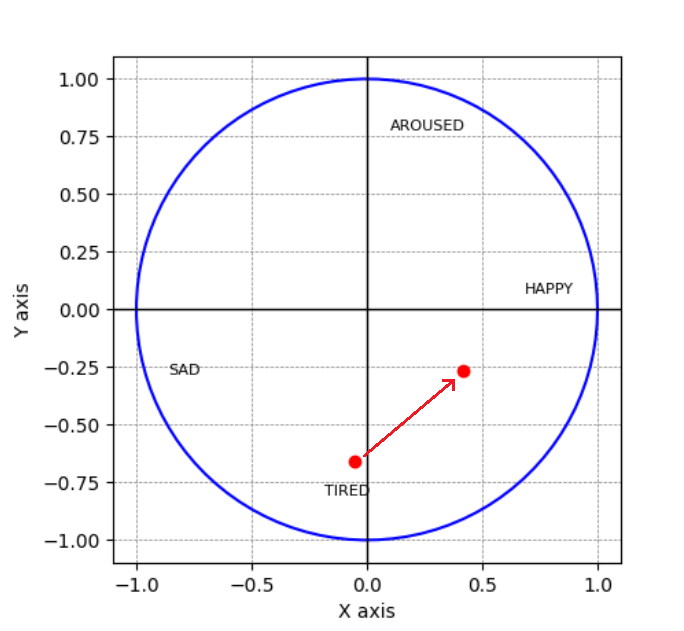}
        \caption{Circumplex Transformation of Melody 105.}
        \label{fig:image1}
    \end{minipage}
    \quad 
    \begin{minipage}[b]{0.4\linewidth}
        \includegraphics[width=\linewidth]{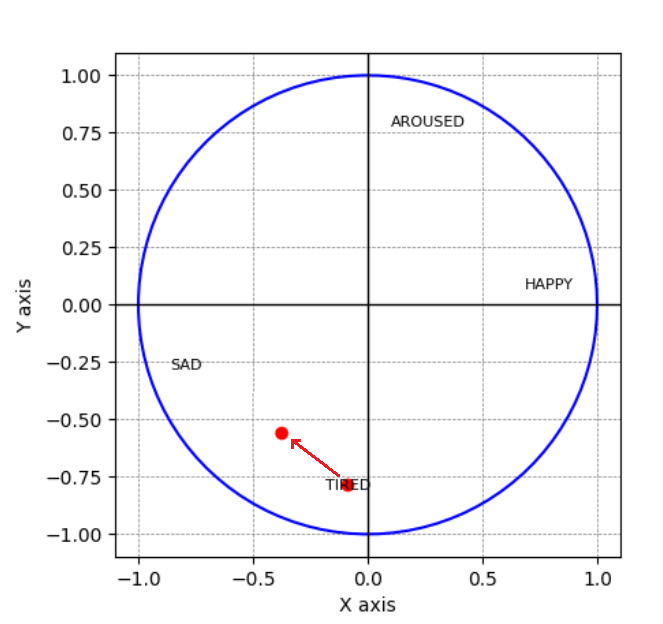}
        \caption{Circumplex Transformation of Melody 076.}
        \label{fig:image2}
    \end{minipage}
\end{figure*}

As shown in Fig. \ref{fig:image1}, it is possible to change the semantic content for a musical piece from happy to sad and the reverse can be seen in Fig. \ref{fig:image2}. These respective shifts were made on melody 105 and 076 from the 4Q Audio Emotion Dataset from the key of B minor to A minor. Samples of the output audio for both transformations are available on our Github page under demos\footnote{\url{https://github.com/aa221/Semantic-Manipulation-of-Music}}. 

We ran this qualitative assessment on melody 076 and melody 105, across four instrument SoundFonts and 108 key transpositions from C0 to B8 representing a testing range of 16.35160 Hz to 7902.133 Hz. This produced a database of 864 entries comprised of each test's probability values before transformations,  probability values after transformations, key, and SoundFont used. We have found that both key and SoundFont significantly impact the transformed melody's emotional content.

\textbf{High level overview:}
One of the SoundFonts selected, "Mario", increased the baseline happiness for every melody to 0.9. This demonstrates that the selection of instrumentation can dramatically impact the emotional content of a given song regardless of key. Despite this, the transposition in key and resultant accompaniment generation still had an impact on the emotional contents of a song. For example, the keys within the C3 to G4 range exhibited a sadder transformation of the song, shifting the happiness probability from its baseline of 0.9 to an average of 0.25. 

\section{Conclusion and Future Work} \label{sec:conclusions}

While we have successfully created an end to end pipeline capable of manipulating the emotional content of music from one emotion to another using an informative way to visualize our song transformations and a deep learning model capable of predicting which Quadrant a given song belongs to, there remains much work to be done.

Our next step in improving this pipeline would be exploring the impact of Accomontage2 on the manipulation of music in isolation. Our analysis has concluded that both keys and Soundfonts have an impact on the shift in emotional content of a given piece of music, but it is difficult to quantify exactly how much impact is coming from each component in our tool chain. 

A second avenue of research we are pursuing is to determine if transposition of the music is affected by distance from the original key as the musical scale operates in a 12-key cycle in each octave. By the 12th step, we will have returned to the original key plus one octave and we posit that the maximal point of difference is approximately six keys away from the original and that the closer to the original key we are at the less the emotion will shift. These possibilities require further testing to verify as we need to study this over multiple songs, keys, and artists to account for any bias in our input data.

Our goal was to provide an \emph{initial} pipeline capable of manipulating music's emotional content, so the applied transformations are not entirely sophisticated. One way of improving the current state of the pipeline is to further increase both the complexity and number of transformations applied to the input audio overall. Features such as timbre, tempo and tone are all areas of focus that can be leveraged to further transform a given audio towards a desired semantic state.

We also hope to evaluate the accuracy of our classifier (and others) using human testing to determine their accuracy in comparison to human emotional perception, which must also take into account a user's demographic data as musical preferences are impacted by what a subject has been exposed to in their lifetime. 

Lastly, it may be worth expanding this concept into a larger system for musical artists interested in manipulating the emotional content of their music at any given stage of their composition process. A more mature SMM platform may allow for artists and copyright holders to customize their music on a per-user basis similar to an on-demand cover or remix of a selected song. In this instance, care should be taken that this is performed within the bounds of copyright law, especially when used by those without ownership of the source music.

However, as stated in \cite{Dissanayake2006}, "Since the beginning of human civilization, music has been used as a device to control social behavior, where it has operated as much to promote solidarity within groups as hostility between competing groups." It is well-known that we can manipulate emotions and influence attitude through music, and that this manipulation may be morally questionable if used, for instance, for commercialization or manipulation of an author's work without their express permission. We encourage those building on our work to remain ethical in their applications of SMM.


\end{document}